\documentclass[12pt]{iopart}
\usepackage{epsfig}
\usepackage{graphicx}

\newcommand{\be}{\begin{equation}}
\newcommand{\ee}{\end{equation}}
\newcommand{\bd}{\begin{displaymath}}
\newcommand{\ed}{\end{displaymath}}
\newcommand{\vsp}{\vspace*{3mm}}

\newcommand{\bra}{\langle}
\newcommand{\ket}{\rangle}
\newcommand{\order}{{\cal O}}

\newcommand{\plus}{{\!+\!}}

\newcommand{\sgn}{{\rm sgn}}
\newcommand{\erf}{{\rm erf}}

\newcommand{\bm}{\ensuremath{\mathbf{m}}}
\newcommand{\bq}{\ensuremath{\mathbf{q}}}

\newcommand{\bR}{\ensuremath{\mathbf{R}}}

\newcommand{\bpsi}{{\mbox{\boldmath $\psi$}}}

\newcommand{\bnull}{{\mbox{\boldmath $0$}}}

\newcommand{\room}{\rule[-0.1cm]{0cm}{0.6cm}}

\newcommand{\bigbra}{\left\langle}
\newcommand{\bigket}{\right\rangle}

\newcommand{\bigroom}{\rule[-0.4cm]{0cm}{1.0cm}}
\newcommand{\one}{{\rm 1\hspace*{-1.5mm}I}}
\begin{document}

\title[Non-equilibrium statistical mechanics of Minority Games]{Non-equilibrium statistical mechanics\\
of Minority Games}
\author{
A C C Coolen}
\address{
Department of Mathematics, King's College London\\ The Strand,
London WC2R 2LS, UK }

\begin{abstract}
\noindent In this paper I give a brief introduction to a family of
simple but non-trivial models designed to increase our
understanding of collective processes in markets, the so-called
Minority Games, and their non-equilibrium statistical mathematical
analysis. Since the most commonly studied members of this family
define disordered stochastic processes without detailed balance,
the canonical technique for finding exact solutions is found to be
generating functional analysis a la De Dominicis, as originally
developed in the spin-glass community.
\end{abstract}


\section{Introduction}

The Minority Game (MG) \cite{ChalZhan97}, a variation on the
so-called El-Farol bar problem \cite{Arth94}, was designed to
understand the cooperative phenomena observed in markets. It
describes agents who each make a binary decision at every point in
time, e.g. whether to buy or sell. Profit is made by those who
find themselves in the minority group, i.e. who end up buying when
most wish to sell, or vice versa. The agents take their decisions
on the basis of an arsenal of individual `strategies', from which
each agent aims to select the optimal one (i.e. the strategy that
leads to the largest number of minority decisions). The dynamic
equations of the MG describe the stochastic evolution of these
strategy selections. Each agent wishes to make profit, but the net
effect of his/her trading
 actions is defined
fully in terms of (or relative to) the actions taken by the other
agents. Hence the model has a significant amount of built-in
frustration. Although in its simplest form (as discussed in this
paper) the MG is a crude simplification of real markets, there is
now also a growing number of more realistic MG spin-off models
\cite{Chalweb}.

For the non-equilibrium statistical mechanicist the MG is an
intriguing disordered system. It has been found to exhibit
non-trivial behaviour, e.g. dynamical phase transitions separating
an ergodic from a highly non-ergodic regime, and to pose a
considerable mathematical challenge. Its stochastic equations do
not obey detailed balance, so there is no equilibrium state and
understanding the model requires solving its dynamics. In this
paper I first give an introduction to the properties and the early
 studies of the MG, seen through the eyes of a non-equilibrium
statistical mechanicist. The second part of the paper is a review
of the more recent applications to the MG of the generating
functional analysis methods of De Dominicis \cite{DeDominicis}.
These have led to exact results in the form of closed macroscopic
order parameter equations, and to clarification of a number of
unresolved issues and debates regarding the nature of the
microscopic laws and various previously proposed approximations.

\section{Definitions and properties}

We consider a community of  $N$ agents, labeled $i=1\ldots N$ (the
`market'). At each discrete iteration step $\ell$ each agent $i$
must take a binary decision $s_i(\ell)\in\{-1,1\}$ (e.g. `sell' or
`buy'). The agents make these decisions on the basis of public
information $I(\ell)$ which they are given at stage $\ell$ (state
of the market, political developments, weather forecasts, etc),
which is taken from some finite discrete set
$\Omega=\{I_1,\ldots,I_p\}$ of size $|\Omega|=p$. For future
convenience we will also define the relative size $\alpha=p/N$.
Note that all agents always receive the same information. Profit
is made at iteration step $\ell$ by those agents who find
themselves in the minority group (by those who wish to buy when
most aim to sell, or vice versa), i.e. by those $i$ for which
$s_i(\ell)[\sum_j s_j(\ell)]<0$.

The actual conversion of the observation of $I(\ell)$ into the $N$
trading decisions $\{s_1(\ell),\ldots,s_N(\ell)\}$ is defined via
so-called trading `strategies', which are basically look-up
tables.
 Every agent $i$ has $S$ decision making strategies
 $\bR_{ia}=(R_{ia}^1,\ldots,R_{ia}^{p})\in\{-1,1\}^p$, with
$a\in \{1,\ldots, S\}$. If strategy $a$ is the one used by agent
$i$ at iteration step $\ell$, and the  public information
$I(\ell)$ is found to be $I(\ell)=I_{\mu(\ell)}$, then agent $i$
will locate the appropriate entry in the table $\bR_{ia}$ and take
the decision
\be
s_{i}(\ell)=R_{ia}^{\mu(\ell)} \label{eq:decisions} \ee
 Thus, given knowledge of the
active strategies of all $N$ agents, the responses of all agents
to the arrival of public information are fully deterministic.

What remains in defining the model is giving a recipe for deciding
the choice of decision strategy $\bR_{ia}$ from the arsenal of $S$
possibilities  as a function of time, for all agents. This is done
as follows. The agents aim to select or discover profitable
strategies within their arsenals, by keeping track of the
strategies' performance in the market. A strategy $\bR_{ia}$ is
profitable for agent $i$ at stage  $\ell$ of the game (i.e. it
puts agent $i$ in the minority group) if and only if
$R_{ia}^{\mu(\ell)}=-\sgn[\sum_j s_j(\ell)]$. Hence we can measure
for each agent $i$ the cumulative performance of strategy
$\bR_{ia}$ by a quantity $p_{ia}$, which is updated at every stage
$\ell$ of the game according to
 \be
p_{ia}(\ell+1)=p_{ia}(\ell)- \frac{\eta}{N}
R_{ia}^{\mu(\ell)}\sum_j s_j(\ell) \label{eq:strategies}
 \ee with $\eta>0$ (the scaling
factor $N$ is introduced for future mathematical convenience).
Each agent $i$ now chooses at every stage $\ell$ of the game the
strategy $a_{i\ell}$ which has the best cumulative performance
$p_{ia}(\ell)$ at that point in time.\vsp

In summary, upon insertion of (\ref{eq:decisions}) into
(\ref{eq:strategies}) and given the external information stream
defined by $\{\mu(\ell)\}$, the dynamics of the MG is defined
fully by the following two coupled non-linear equations for the
evolving strategy performance measures $\{p_{ia}\}$ (with
$A(\ell)$ giving the total market bid at stage $\ell$):
 \be
p_{ia}(\ell+1)=p_{ia}(\ell)- \frac{\eta}{N} R_{ia}^{\mu(\ell)}
A(\ell)~~~~~~~A(\ell)= \sum_j R_{j a_j(\ell)}^{\mu(\ell)}
\label{eq:combined}
 \ee
 \be
a_i(\ell)={\rm arg}~{\rm max}_{a\in\{1,\ldots,S\}} p_{ia}(\ell)
\label{eq:evolution} \ee In the original model \cite{ChalZhan97}
the information $\mu(\ell)$ was defined as a deterministic
function of the macroscopic market history, i.e. of the terms
$A(\ell^\prime)=\sum_j s_j(\ell^\prime)=\sum_{j R_j
a_j(\ell^\prime)}^{\mu(\ell^\prime)}$ with $\ell^\prime<\ell$:
$\mu(\ell)=K[A(\ell-1),A(\ell-2),\ldots]$, for some function
$K[\ldots]$. In this formulation the process
(\ref{eq:combined},\ref{eq:evolution}) is fully deterministic but
non-local in time. Furthermore, there is quenched disorder (the
realization of the look-up tables $\{\bR_i\}$) and a large degree
of built-in competition and frustration. \vsp

\begin{figure}[t]
\vspace*{3mm}\hspace*{15mm}
 \setlength{\unitlength}{0.85mm}
\begin{picture}(100,100)
\put(20,20){\epsfxsize=95\unitlength\epsfbox{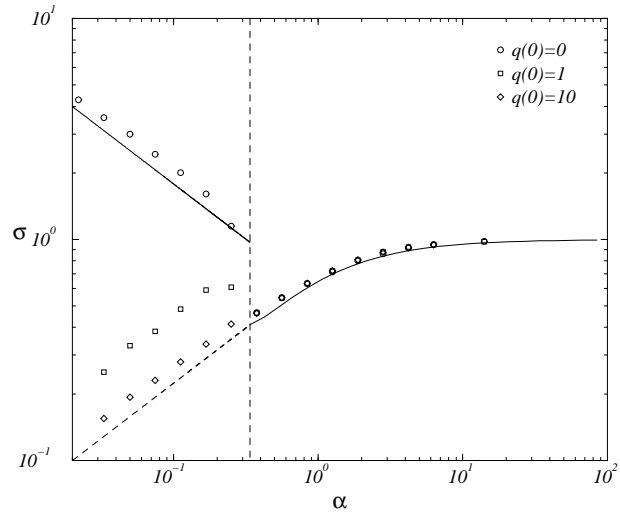}}
\end{picture}
\vspace*{-15mm} \caption{Dependence of the volatility $\sigma$ on
the re-scaled size $\alpha=p/N$ of the external information set
$\Omega$. The markers refer to numerical simulations following
different initial conditions (whose details will be explained
later), with $S=2$ (two strategies per agent) and $N=4000$. The
value $\sigma=1$ would have been obtained for purely random
decision making.} \label{fig:volatility}
\end{figure}

The main quantity of interest to those caring about real markets
is the so-called volatility, the measure of macroscopic market
fluctuations. In the present model the appropriate object is the
re-scaled and time-averaged variance $\sigma^2$ of the total bid
$A(\ell)$ (since the time-averaged expectation value of the total
bid is zero):
\be
\sigma^2=\lim_{\tau\to\infty}\frac{1}{\tau N}\sum_{\ell=1}^\tau
A^2(\ell) \label{eq:volatility} \ee This quantity was found in
simulations to depend in a nontrivial way on the relative size
$\alpha$ of the set $\Omega$, similar to e.g. Figure
\ref{fig:volatility}, which seemed to suggest to agents the
possibility of market prediction (since random decision making
corresponds to $\sigma=1$) and to theorists the presence of a
non-equilibrium phase transition. Since simulations also indicated
that the behaviour of the volatility did not depend qualitatively
on the value of $S$ (the number of strategies per agent, provided
this number remains finite as $N\to \infty$), most of the research
has concentrated on the simplest nontrivial case $S=2$. Here it is
possible to reduce the dimensionality of the problem further upon
defining \be q_i=\frac{p_{i1}-p_{i2}}{2},~~~~~~
\xi_i^\mu=\frac{R_{i1}^\mu-R_{i2}^\mu}{2},~~~~~~
\Omega_\mu=\frac{\sum_j[R_{j1}^\mu+R_{j2}^\mu]}{2\sqrt{N}}
\label{eq:disorder} \ee We can now write the process
(\ref{eq:combined},\ref{eq:evolution}) in the deceivingly compact
form
 \be
q_i(\ell+1)=q_i(\ell)-\frac{\eta}{\sqrt{N}}~\xi_i^{\mu(\ell)}
\left[\Omega_{\mu(\ell)}+\frac{1}{\sqrt{N}}\sum_j\xi_j^{\mu(\ell)}{\rm
sgn}[q_j(\ell)]\right] \label{eq:S2dynamics} \ee which in the
standard MG is non-local in time due to the  dependence of
$\mu(\ell)=K[A(\ell-1),A(\ell-2),\ldots]$ (with
$A(\ell^\prime)=\sum_j {\rm sgn}[q_j(\ell^\prime) ]$) on the past
of the system.

\section{Early studies and generalizations}

The first major step forward after the MG's introduction in
\cite{ChalZhan97} was \cite{Cavagna99}, where it was shown that
the volatility $\sigma$ remains largely unchanged if, instead of
being calculated from the true market history, the variables
$\mu(\ell)$ are simply drawn at random from $\{1,\ldots,\alpha
N\}$ (at every iteration step $\ell$, independently and with equal
probabilities). Apparently the agents are not predicting the
market: the observed phenomena are the result of complex
collective processes, which depend only on all agents responding
to the {\em same} information (whether correct or nonsensical).
The mathematical consequences of having the numbers $\mu(\ell)$
drawn at random, however, are profound: the equations
(\ref{eq:S2dynamics}) now describe a disordered Markov process
(without detailed balance) and analysis comes within reach. In
\cite{SaviManuRiol99} it was shown that the relevant quantities
studied in the MG can be scaled such that they become independent
of the number of agents $N$ when this number becomes very large.
 An interesting
generalization of the game was the introduction of agents'
decision noise \cite{CavaGarrGiarSher99}, which was shown not only
to improve worse than random behaviour but also, more
surprisingly, to be able to make it better than
random\footnote{Using a phenomenological theory for the
volatility, based on so-called `crowd-anticrowd' cancellations 
this effect was partially explained in
\cite{Johnson2,Johnson3}.}. In \cite{ChalletMarsili99} one finds
the first observation in the MG of the  so-called `frozen agents'.
These are agents for which asymptotically $q_i(\ell)\sim \ell$
($\ell\to\infty$), in the language of (\ref{eq:S2dynamics}), and
who in view of the definition $q_i=\frac{1}{2}[p_{i1}-p_{i2}]$
will eventually end up always playing the same strategy. The four
papers
\cite{Cavagna99,SaviManuRiol99,CavaGarrGiarSher99,ChalletMarsili99}
more or less paved the way for statistical mechanical theory.

The first attempt at solution of the MG is found in
\cite{CMZ2000,MCZ2000}, whose authors studied the MG with decision
noise and argued that for $N\to\infty$, and after appropriate
temporal course-graining, one is allowed to neglect the
fluctuations in $\sgn[q_i(\ell)+z_i(\ell)]$ and concentrate on the
expectation values $m_i(\ell)=\int\!dz~P(z) \sgn[q_i(\ell)+z]$. In
terms of the re-scaled time $t=\ell/N$ the authors then derived
deterministic equations in the limit $N\to\infty$, of the form
$dm_i(t)/dt=f_i[\bm(t)]$, which were shown to have a Lyapunov
function. This allowed for a ground state analysis using
equilibrium statistical mechanics, in combination with standard
replica theory \cite{Mezardetal} to carry out the disorder average
 (based on the introduction of a fictitious temperature which is
sent to zero {\em after} having taken the limits $N\to \infty$ and
$n\to 0$, where $n$ denotes the replica dimension, similar to e.g.
\cite{Kuhn}). The authors of \cite{CMZ2000,MCZ2000} found a phase
transition at $\alpha\downarrow \alpha_c\approx 0.33740$ (in
agreement with simulation evidence, see e.g. Figure
\ref{fig:volatility}), but ran into mathematical difficulties for
$\alpha<\alpha_c$.

The authors of \cite{CMZ2000,MCZ2000} then initiated a discussion
\cite{CMZPRL,CGGSPRL} about the validity of the assumptions in
\cite{CMZ2000,MCZ2000} and the reliability of the simulations in
\cite{CavaGarrGiarSher99}. Following a further study
\cite{GMS2000}, in which a Fokker-Planck equation was derived for
the MG (and solved numerically), and where fluctuations were
claimed not to be negligible, this discussion appeared to be put
to rest in \cite{MC2001}  with the conclusions that the
simulations of \cite{CavaGarrGiarSher99} had not yet equilibrated
(acknowledged already in \cite{CGGSPRL}), and that the microscopic
fluctuations in the MG can not be neglected after all (except
perhaps in special cases). It should be noted, however, that the
microscopic definitions of
\cite{CavaGarrGiarSher99,CGGSPRL,GMS2000} are different from those
of \cite{CMZ2000,CMZPRL,MC2001}. This state of affairs was
unsatisfactory: there was still no exact solution of the MG
(whether for statics or dynamics), the status of the assumptions
and approximations made by the various authors remained unclear,
in both \cite{GMS2000} and \cite{MC2001} Fokker-Planck equations
are derived for the MG, but with different diffusion matrices, and
(finally) there were conflicting statements on the differences
between multiplicative and additive decision noise with regard to
the volatility. Clearly, since the MG defines a disordered
stochastic process without detailed balance, the natural
analytical approach is to study its dynamics. In the remainder of
this paper I show how the generating functional techniques
developed by De Dominicis \cite{DeDominicis} can be used to solve
various versions of the MG analytically. Full details of this
analysis can be found in the trio
\cite{HeimelCoolen2001,CoolenHeimelSherr2001,CoolenHeimel2001}.

\section{Solution of the Batch MG}

In order to circumvent the debates relating to the temporal course
graining of the MG process and concentrate fully on the disorder,
the first generating functional analysis study of the MG
\cite{HeimelCoolen2001} was carried out for a modified so-called
`batch' version of the model. Instead of the equations
 (\ref{eq:S2dynamics}) (where state changes follow each randomly drawn
 information index $\mu(\ell)$), in the batch MG the dynamics is
 defined directly in terms of an average over all possible
 choices $\mu\in\{1,\ldots,\alpha N\}$ for the external information:
 \bd
q_i(\ell+1)=q_i(\ell)-\frac{1}{\alpha N }\sum_{\mu=1}^{\alpha N}
\xi_i^{\mu}
\left\{\Omega_{\mu}\plus\frac{1}{\sqrt{N}}\sum_j\xi_j^{\mu}{\rm
sgn}[q_j(\ell)]\right\}+\theta_i(\ell)\ed $~$\\[-10mm]
\be
\label{eq:S2batchdynamics} \ee Here we have also chosen
$\eta=\sqrt{N}$ (to ensure $\order(N^0)$ characteristic
time-scales for the batch version of the MG). This defines a
disordered deterministic map, since the stochasticity has now been
removed. An external force $\theta_i(\ell)$ was added in order to
define response functions later.

Generating functional analysis a la De Dominicis
\cite{DeDominicis} is based on evaluation of the following
disorder-averaged\footnote{Disorder averaging will be denoted as
$\overline{[\ldots]}$. The disorder is in the variables
$\xi_i^\mu$ and $\Omega_\mu$, see the definitions
(\ref{eq:disorder}), with $\mu\in\{1,\ldots,\alpha N\}$ and with
each of the $2\alpha N^2$ strategy table entries $R_{ia}^\mu$
drawn independently at random from $\{-1,1\}$ with equal
probabilities. } generating functional: \be
\overline{Z\left[\bpsi\right]}= \overline{\room\bigbra
e^{-i\sum_{i=1}^N\sum_{t\geq 0}~\psi_i(t)q_i(t)}\bigket}
\label{eq:generating_functional} \ee From this object one can
obtain the main disorder-averaged quantities of interest, such as
individual averages, multiple-time covariances, and response
functions, e.g.
\be
\overline{\bra q_i(\ell)\ket}=\lim_{\bpsi\to
\bnull}\frac{i\partial
\overline{Z\left[\bpsi\right]}}{\partial\psi_i(\ell)} \ee
\be
\overline{\bra q_i(\ell)q_j(\ell^\prime)\ket}=-\lim_{\bpsi\to
\bnull}\frac{\partial^2
\overline{Z\left[\bpsi\right]}}{\partial\psi_i(\ell)\partial\psi_j(\ell^\prime)}
\ee
\be
\frac{\partial\overline{\bra
q_i(\ell)\ket}}{\partial\theta_j(\ell^\prime)}=\lim_{\bpsi\to
\bnull}\frac{i\partial^2
\overline{Z\left[\bpsi\right]}}{\partial\psi_i(\ell)\partial\theta_j(\ell^\prime)}
\ee One next defines suitable (time-dependent) fields which must
depend linearly on an extensive number of the statistically
independent disorder variables $\{R_{ia}^\mu\}$, and which drive
the dynamics (in spin-glass models these would have been the
actual local magnetic fields), such as
 $x^\mu(\ell)=\sqrt{2/N}\sum_i {\rm sgn}[q_i(\ell)]\xi^\mu_i$.
 One then writes  (\ref{eq:generating_functional}) as a sum
(or integral) over all possible paths taken by these fields. In
practice this is done via the insertion into
(\ref{eq:generating_functional}) of integrals over appropriate
$\delta$-distributions, e.g. of
\be
1=\prod_{\mu=1}^{\alpha N}\prod_{\ell\geq 0}\left\{\int\!
dx^\mu(\ell)~\delta\left[ x^\mu(\ell)-\frac{\sqrt{2}}{
  \sqrt{N}}\sum_i {\rm sgn}[q_i(\ell)]\xi^\mu_i
\right]\right\} \label{eq:insert_delta}
 \ee This procedure is carried out until all
disorder variables have been concentrated into such fields; here
this requires further fields in addition to the $x^\mu(\ell)$.
Upon writing the $\delta$-functions in expressions such as
(\ref{eq:insert_delta}) in integral form, the independent disorder
variables $\{R_{ia}^\mu\}$ will all occur linearly in exponents,
so that the disorder average can be carried out. At the end one
finds an expression involving an integral over all possible values
of the single-site correlation and response functions
\begin{eqnarray}
C_{tt^\prime}&=&\frac{1}{N}\sum_{i=1}^N \overline{\bra
\sgn[q_i(t)]\sgn[q_i(t^\prime)]\ket} \label{eq:defineC}\\
G_{tt^\prime}&=&\frac{1}{N}\sum_{i=1}^N
\frac{\partial}{\partial\theta_i(t^\prime)}\overline{\bra
\sgn[q_i(t)]\ket} \label{eq:defineG}\end{eqnarray}
 and a further set of related  order
parameter kernels
$\{\hat{C}_{tt^\prime},\hat{G}_{tt^\prime},L_{tt^\prime},\hat{L}_{tt^\prime}\}$:
 \be
 \overline{Z\left[\bpsi\right]}=\int\!{\cal D}C{\cal
 D}\hat{C}{\cal D}G{\cal D}\hat{G}{\cal D}L {\cal D}\hat{L}
 ~e^{N\Psi[C,\hat{C},G,\hat{G},L,\hat{L}]}
 \label{eq:steepest_descent}
 \ee
with the abbreviation ${\cal D}C\equiv \prod_{tt^\prime\geq
0}dC_{tt^\prime}$, and with $\Psi[\ldots]=\order(N^0)$. In the
limit $N\to\infty$ the integral in (\ref{eq:steepest_descent}) is
evaluated by steepest descent, giving closed dynamical equations
for the kernels $\{C,\hat{C},G,\hat{G},L,\hat{L}\}$, from which
$\{\hat{C},\hat{G},L,\hat{L}\}$ can be eliminated. This leaves a
theory described by closed equations for $C$ and $G$ only, which
turns out to describe an effective stochastic `single agent'
process of the form
 \be
q(t+1)=q(t)+\theta(t)-\alpha \sum_{t^\prime\leq t}[\one+G]^{-1}_{t
t^\prime}~\sgn[q(t^\prime)]+\sqrt{\alpha}~\eta(t)
\label{eq:single_agent}
 \ee In (\ref{eq:single_agent}) one has a retarded self-interaction and a
zero-average Gaussian noise $\eta(t)$ with
 \be
 \bra\eta(t)\eta(t^\prime)\ket=\left[(\one+ G)^{-1}D(\one+ G^\dag)^{-1}\right]_{t
t^\prime} ~~~~~~~~ D_{tt^\prime}=1+ C_{tt^\prime}
\label{eq:covariances} \ee The equations from which to solve the
two kernels $\{C,G\}$ self-consistently are defined in terms of
the non-Markovian single-agent process
(\ref{eq:single_agent},\ref{eq:covariances}) as follows: for all
$t,t^\prime\in {\rm I\!N}$
 \be
 C_{tt^\prime}=\bra
\sgn[q(t)]\sgn[q(t^\prime)]\ket,~~~~~
 G_{tt^\prime}=
\frac{\partial}{\partial\theta(t^\prime)} \bra\sgn[q(t)]\ket
\label{eq:CG_equations} \ee
 Finding the general solution of
(\ref{eq:CG_equations}) analytically is extremely hard; one has to
restrict oneself in practice to working out the first few time
steps, to stationary solutions, or to numerical solution.

\begin{figure}[t]
\vspace*{-7mm}
\hspace*{10mm}
 \setlength{\unitlength}{0.85mm}
\begin{picture}(100,100)
\put(0,20){\epsfxsize=80\unitlength\epsfbox{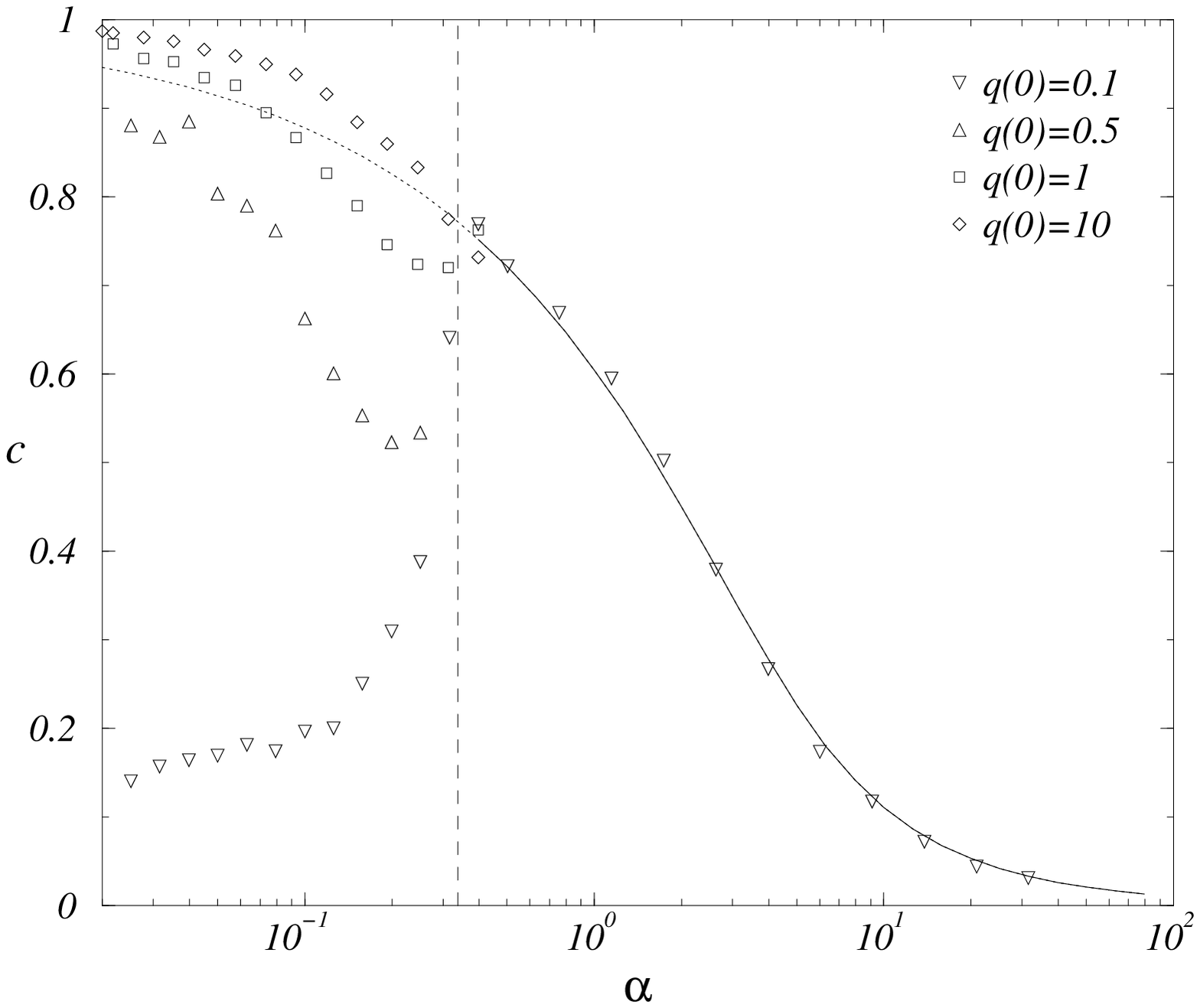}}
\put(85,20){\epsfxsize=80\unitlength\epsfbox{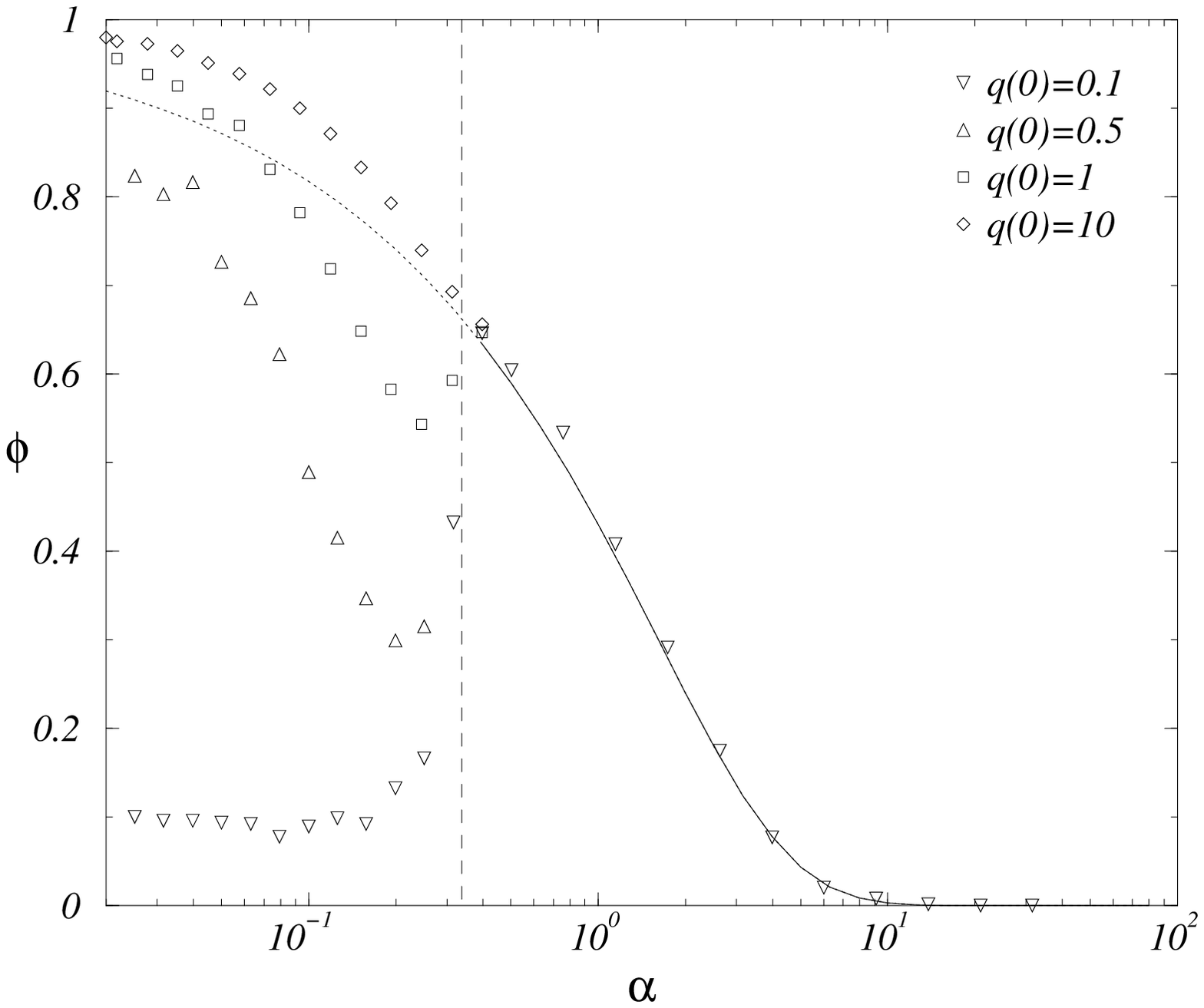}}
\end{picture}
\vspace*{-15mm} \caption{The persistent covariance
$c=\lim_{t\to\infty}C(t)$ (left picture) and the fraction $\phi$
of frozen agents (right picture) in the stationary state. The
markers are obtained from individual simulation runs, with
$N=4000$ and various homogeneous initial conditions (where
$q_i(0)=q(0)$ $\forall i$), measured after 1000 iteration steps.
The solid curves to the right of the critical point (vertical
dashed line in each picture) are the theoretical predictions,
given by the solution of
 (\ref{eq:c},\ref{eq:phi}).
The dotted curves to the left of the critical point are their
continuations into the $\alpha<\alpha_c$ regime (where they should
no longer be correct).} \label{fig:c_phi}
\end{figure}

If, for instance, one investigates asymptotic stationary solutions
of the form $C_{tt^\prime}=C(t-t^\prime)$ and
$G_{tt^\prime}=G(t-t^\prime)$, and furthermore assumes absence of
anomalous response (i.e. a finite value for the integrated
response $\chi=\sum_{t>0}G(t)$) it is not difficult
\cite{HeimelCoolen2001} to obtain an exact closed equation for the
persistent correlation $c=\lim_{t\to \infty}C(t)$:
\begin{equation}
  c=1-(1- \frac{1+c}{\alpha})~\erf\left[\sqrt{\frac{\alpha}{2(1+ c)}}\right]
 -\sqrt{\frac{2(1+ c)}{\pi \alpha}}e^{-\frac{\alpha}{2(1+c)}}.
  \label{eq:c}
\end{equation}
from which $c$ follows as a function of the parameter $\alpha$.
Similarly one can find exact expressions for the fraction $\phi$
of `frozen' agents  and for the integrated response $\chi$:
\begin{eqnarray}
\phi&=&1-\erf[\sqrt{\alpha/2(1+c)}] \label{eq:phi} \\
\chi&=&(1-\phi)/(\alpha-1+\phi) \label{eq:chi} \end{eqnarray} For
$\alpha\to\infty$ one finds
$\lim_{\alpha\to\infty}c=\lim_{\alpha\to\infty}\phi=\lim_{\alpha\to\infty}\chi=0$.
 It
follows from (\ref{eq:chi}) that the assumption $\chi<\infty$
underlying (\ref{eq:c},\ref{eq:phi},\ref{eq:chi}) breaks down when
$\alpha=1-\phi$. This defines a critical value $\alpha_c$
signaling the onset of non-ergodicity, which can be calculated in
combination with (\ref{eq:phi}), giving $\alpha_c\approx 0.33740$
(the same value obtained earlier in \cite{CMZ2000,MCZ2000}). The
results of solving (\ref{eq:c}) and (\ref{eq:phi}) numerically, as
a function of $\alpha$, are shown in Figure \ref{fig:c_phi}. One
observes perfect agreement with simulations for $\alpha>\alpha_c$,
and, as expected, disagreement and ergodicity breaking for
$\alpha<\alpha_c$ (where the condition $\chi<\infty$ is violated).
The volatility is found not to be a natural order parameter of the
MG (nor expressible directly in terms of the solution of
(\ref{eq:c},\ref{eq:phi}), not even for $\alpha>\alpha_c$; its
calculation required approximations, resulting in Figure
\ref{fig:volatility} (shown together with simulation results of
the same specifications as those in Figure \ref{fig:c_phi}).

\begin{figure}[t]
\vspace*{-10mm}\hspace*{15mm}
 \setlength{\unitlength}{0.9mm}
\begin{picture}(100,100)
\put(20,20){\epsfxsize=95\unitlength\epsfbox{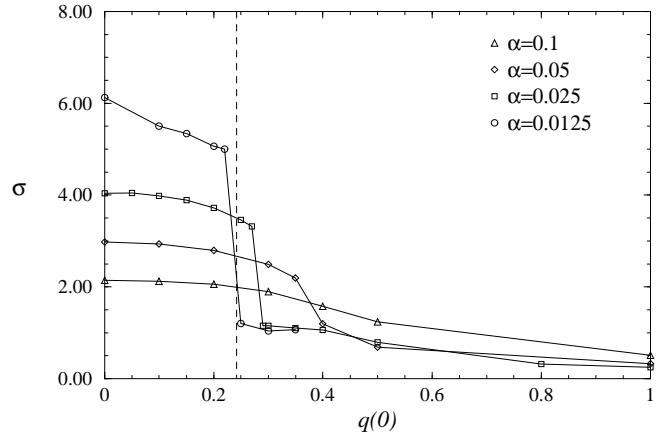}}
\end{picture}
\vspace*{-15mm} \caption{Evidence for the existence of a critical
 initial strategy valuation $q_c(0)$ below which a high-volatility
solution $\sigma\sim\alpha^{-1/2}$ $(\alpha\to 0)$ exists.
Connected markers show the results of
 individual simulations, with $N=4000$ and homogeneous initial
conditions $q_i(0)=q(0)$ $\forall i$, measured after 1000
iteration steps. The data are compatible with the  prediction
$q_c(0)\approx 0.242$ (dashed line).} \label{fig:q_crit}
\end{figure}

It should be emphasized that for $\alpha<\alpha_c$ the macroscopic
equations
(\ref{eq:single_agent},\ref{eq:covariances},\ref{eq:CG_equations})
are still exact; finding their stationary solutions analytically,
however, is much more difficult. Some of the results which have
been obtained for the $\alpha<\alpha_c$ regime, for instance, are
scaling properties of solutions for $\alpha\to 0$ (shown in Figure
\ref{fig:volatility}). Low volatility solutions $\sigma\sim
\alpha^{\frac{1}{2}}$ $(\alpha\to 0$) are found following initial
conditions with large $|q_i(0)|$, whereas high-volatility
solutions with $\sigma\sim \alpha^{-\frac{1}{2}}$ $(\alpha\to 0$)
are found following initial conditions with small $|q_i(0)|$. Upon
making some simple approximations one can also calculate an
expression for the critical initial value for the $|q_i(0)|$,
$q_c(0)\approx 0.242$, bordering the region of existence of the
high-volatility solution, see Figure \ref{fig:q_crit}.

\section{Solution of the Batch MG with decision noise}

The Batch MG with decision noise is obtained upon making in
(\ref{eq:S2batchdynamics}) the replacement $\sgn[q_i(t)]\to
\sigma[q_i(\ell),z_i(\ell)|T_i]$, with the $\{z_i(\ell)\}$
representing zero-average and unit-variance identically
distributed and independent random variables. The parameters
$T_i\geq 0$ measure the noise levels of the individual agents. The
function $\sigma[\ldots|T]$ obeys $\sigma[q,z|T]\in\{-1,1\}$,
$\sigma[q,z|0]=\sgn[q]$ (for $T\to 0$ we return to the previous
model) and $\int\!dz~P(z)\sigma[q,z|\infty]=0$ ($T\to\infty$
implies purely random decisions). Examples are
\begin{eqnarray} {\rm additive~noise:}&~~~~~&
\sigma[q,z|T]=\sgn[q+Tz] \label{eq:additive}\\{\rm
multiplicative~noise:}&~~~~~&\sigma[q,z|T]=\sgn[q]~\sgn[1+Tz]
\label{eq:multiplicative}
\end{eqnarray} The deterministic
equations (\ref{eq:S2batchdynamics}) are thus replaced by the
following stochastic ones:
 \bd
 \hspace*{-10mm}
q_i(\ell+1)=q_i(\ell)-\frac{1}{\alpha N}\sum_{\mu=1}^{\alpha N}
\xi_i^{\mu}
\left\{\Omega_{\mu}+\frac{1}{\sqrt{N}}\sum_j\xi_j^{\mu}~\sigma[q_j(\ell),z_j(\ell)|T_j]\right\}+\theta_i(\ell)\ed
$~$\\[-10mm]
\be
\label{eq:S2noisybatchdynamics} \ee The analytical procedure
outlined for the previous model (\ref{eq:S2batchdynamics}) is
adapted quite easily to its generalization
(\ref{eq:S2noisybatchdynamics}). Again one finds, in the limit
$N\to\infty$,  closed equations for the disorder-averaged
correlation and response functions
(\ref{eq:defineC},\ref{eq:defineG}), describing an effective
stochastic `single agent' process:
 \bd
q(t+1)=q(t)+\theta(t)-\alpha \sum_{t^\prime\leq t}[\one+G]^{-1}_{t
t^\prime}\sigma[q(t^\prime),z(t^\prime)|T]+\sqrt{\alpha}~\eta(t)
\ed $~$\\[-10mm]
\be
\label{eq:noisy_single_agent}
 \ee
(parametrized by a noise strength $T$, so that we must denote
averages henceforth as $\bra \ldots\ket_T$). The covariances of
the zero-average Gaussian noise $\eta(t)$ are still given by
(\ref{eq:covariances}); the independent random variables
$\{z(t)\}$ are distributed exactly as the site-variables
$\{z_i(\ell)\}$ in (\ref{eq:S2noisybatchdynamics}). The order
parameter equations (\ref{eq:CG_equations}) are now found to be
replaced by
\begin{eqnarray}
 C_{tt^\prime}&=&\int_0^\infty\!dT~W(T)~\bra
\sgn[q(t)]\sgn[q(t^\prime)]\ket_T \label{eq:noisyC}\\
 G_{tt^\prime}&=&
\int_0^\infty\!dT~W(T)~ \frac{\partial}{\partial\theta(t^\prime)}
\bra\sgn[q(t)]\ket_T \label{eq:noisyG}
\end{eqnarray}
with the distribution of noise strengths
\be
W(T)=\lim_{N\to\infty}\frac{1}{N}\sum_{i=1}^N \delta[T-T_i] \ee
Making the choice $W(T)=\delta(T)$ (i.e. deterministic decision
making) immediately brings us back to
(\ref{eq:single_agent},\ref{eq:covariances},\ref{eq:CG_equations}),
as it should.

\begin{figure}[t]
\vspace*{4mm}\hspace*{15mm}
 \setlength{\unitlength}{0.85mm}
\begin{picture}(100,100)
\put(20,20){\epsfxsize=90\unitlength\epsfbox{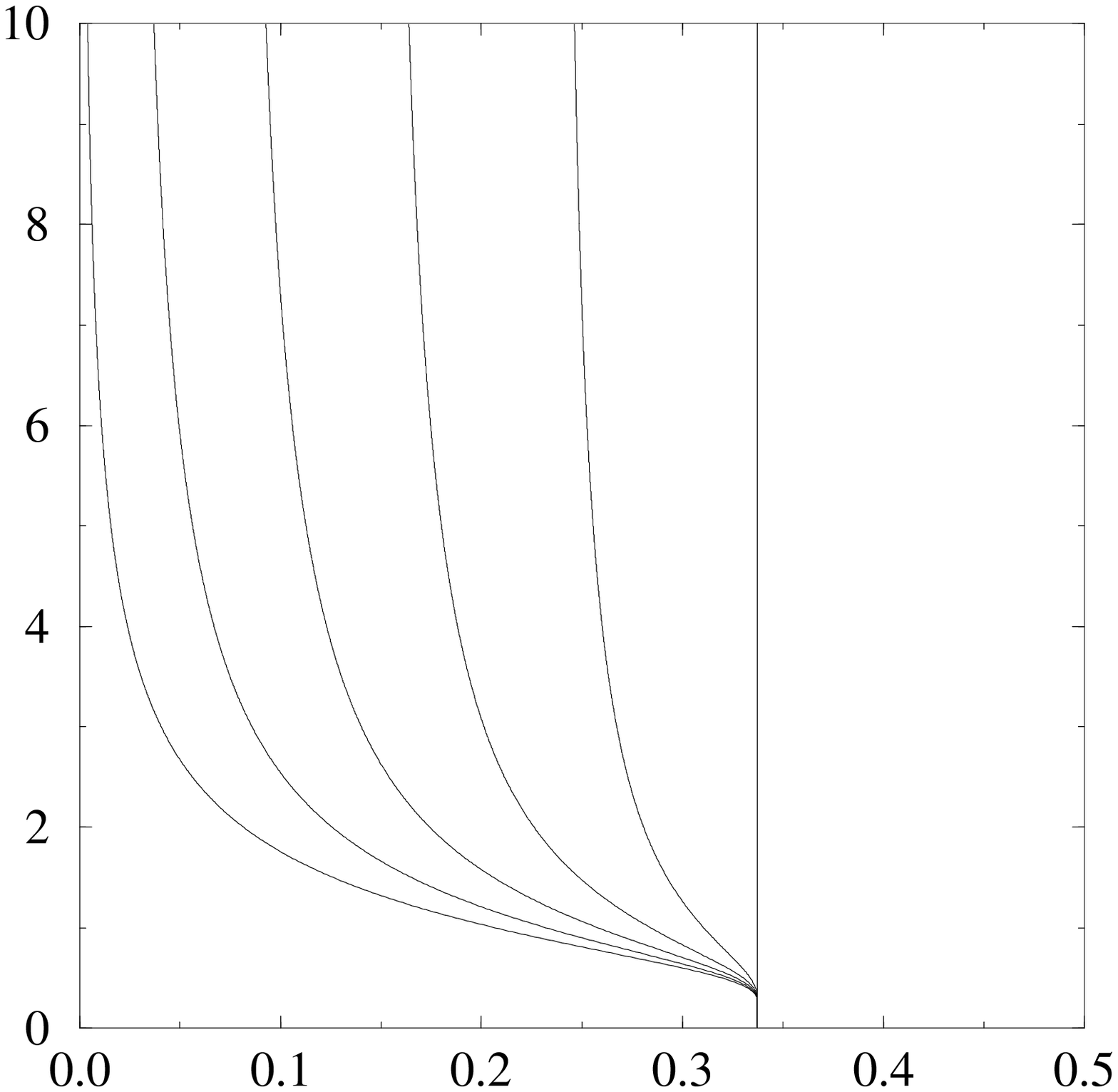}}
\put(66,25){$\alpha$} \put(23,67){$T$}
\end{picture}
\vspace*{-17mm} \caption{Transition lines for the Batch MG with
multiplicative decision noise (\ref{eq:multiplicative}),
$P(z)=(2\pi)^{-1/2}e^{-z^2/2}$, and
$W(T^\prime)=\epsilon\delta[T^\prime-T]+(1-\epsilon)\delta[T^\prime]$
with $\epsilon\in\{0,0.2,0.4,0.6,0.8,1\}$
 (from right to left).
Each line is defined by the condition
$\chi\equiv\sum_{t>0}G(t)=\infty$ and separates a non-ergodic
phase (to the left) from an ergodic phase (to the right). For
additive noise our theory predicts the $T$-independent (vertical)
line, i.e. the $\epsilon=0$ curve, for any choice of $W(T)$. }
\label{fig:phasediags}
\end{figure}

In the present model one can as before calculate stationary
solutions without anomalous response, and calculate order
parameters such as $c$ and $\phi$ as a function of $\alpha$ and
the distribution $W(T)$, and for different types of decision
noise.  A phase transition to a highly non-ergodic state will
again be marked by $\chi\equiv\sum_{t>0}G(t)=\infty$. Here I
restrict myself to showing phase diagrams for some simple choices,
see Figure \ref{fig:phasediags}. The theory reveals that, in the
ergodic regime, for additive decision noise (\ref{eq:additive})
the location of the $\chi=\infty$ transition is completely
independent of $W(T)$\footnote{The explanation for this seems to
be that the $\chi=\infty$ transition is determined purely by the
properties of the `frozen agents'. These agents, for which
$q_i(\ell)\sim \ell$ as $\ell\to \infty$ will asymtotically always
overrule the noise term in (\ref{eq:additive}). In contrast, for
multiplicative noise (\ref{eq:multiplicative}) the noise will
retain the potential to change decisions, even for `frozen'
agents.}; the same is true for
 the values of the order parameters $c$ and $\phi$. Calculating the solution of equations
(\ref{eq:noisy_single_agent},\ref{eq:covariances},\ref{eq:noisyC},\ref{eq:noisyG})
for the first few time-steps \cite{CoolenHeimelSherr2001} also
reveals the mechanism of how the introduction of decision  noise
can indeed have the counter-intuitive effect (observed already in
\cite{CavaGarrGiarSher99}) of reducing the market volatility
$\sigma$, by damping global oscillations.

\section{Solution of the On-line Minority Game}

Having carried out the generating functional analysis of the two
previous batch versions of the MG (with $S=2$, and with
deterministic or noisy decision making), the obviouys next step is
to return to the original (more complicated) dynamics
(\ref{eq:S2dynamics}) where state changes followed individual
presentations of external information, but now generalized to
include decision noise:
 \bd
 \hspace*{-10mm}
q_i(\ell+1)=q_i(\ell)-\frac{\eta}{\sqrt{N}}~\xi_i^{\mu(\ell)}
\left\{\Omega_{\mu(\ell)}+\frac{1}{\sqrt{N}}\sum_j\xi_j^{\mu(\ell)}\sigma[q_j(\ell),z_j(\ell)]\right\}
+\theta_i(\ell) \ed $~$\\[-10mm]
\be
\label{eq:S2onlinedynamics} \ee The $\mu(\ell)$ are drawn randomly
and independently from $\{1,\ldots,\alpha N\}$, introducing a
second element of stochasticity on top of the decision making
noise variables $\{z_j(\ell)\}$. However, the individual (noisy)
updates are now small ($\order(N^{-\frac{1}{2}})$, so for
$N\to\infty$ we may attempt a continuous-time description. This
switch to continuous time (in particular: whether upon doing so
the fluctuations can be taken as Gaussian, or even be discarged)
was the subject of the discussion in \cite{CMZPRL,CGGSPRL} and
subsequent papers.

The temporal regularization problem can be resolved  using a
procedure described in \cite{Bedeaux}. The idea is to link the
process (\ref{eq:S2onlinedynamics})  to a continuous time one by
defining random durations for each of the iterations in
(\ref{eq:S2onlinedynamics}), distributed according to \bd {\rm
Prob}\left[\ell~{\rm steps~at~time}~t\in {\rm
I\!R}^+\right]=\frac{1}{\ell
!}\left[\frac{t}{\Delta_N}\right]^\ell e^{-t/\Delta_N} \ed The
parameter $\Delta_N$ gives the average real-time duration of a
single  iteration of (\ref{eq:S2onlinedynamics}). Upon writing the
Markov process (\ref{eq:S2onlinedynamics}) in probabilistic
form\footnote{We leave aside the external fields for now, which
only serve to define response functions.},
 \be p_{\ell+1}(\bq)=\int\!d\bq^\prime~W(\bq|\bq^\prime)
p_\ell(\bq^\prime) \label{eq:markovchain} \ee
we find the new
continuous-time process to be defined as
\be
\frac{d}{dt}p_t(\bq)=\frac{1}{\Delta_N}\left\{\bigroom
\int\!d\bq^\prime~W(\bq|\bq^\prime)p_t(\bq^\prime)-p_t(\bq)\right\}
\label{eq:masterequation} \ee The price paid is  uncertainty about
where one is on the time axis. This uncertainty, however, will
vanish if we choose $\Delta_N\to 0$ for $N\to\infty$. Upon
studying the scaling properties with $N$ of the process
(\ref{eq:masterequation}) via the Kramers-Moyal expansion, for the
 kernel $W(\bq|\bq^\prime)$ corresponding to
(\ref{eq:S2onlinedynamics}), one can establish the following
facts:
\begin{itemize}
\item
The correct scaling for the parameters $\Delta_N$ and $\eta$ is:
$\Delta_N=\order(N^{-1})$, $\eta=\order(N^0)$
\item As $N\to\infty$ one obtains a Fokker-Planck equation for
 individual components $q_i$: \bd
\frac{d}{dt}p_t(q_i)=\frac{\partial}{\partial
q_i}\left[p_t(q_i)F_i(q_i,t)\right]+\frac{1}{2}\frac{\partial^2}{\partial
q_i^2}\left[p_t(q_i)D_i(q_i,t)\right]+\order(\frac{1}{\sqrt{N}})
\ed \vspace*{-1mm}
\item The process for
$\bq=(q_1,\ldots,q_N)$, however,  cannot for $N\to\infty$ be
described by a Fokker-Planck equation (non-Gaussian fluctuations
in the $q_i$ cannot a priori be discarged; although small, they
might conspire for an $N$-component system).
\end{itemize}
Working out the diffusion matrix $D_{ij}(\bq)$ of the full
process, for additive decision noise (\ref{eq:additive}), for
comparison with previous results in \cite{GMS2000} and
\cite{MC2001} (where the MG  was claimed to obey a Fokker-Planck
equation), reveals that $D_{ij}(\bq)=D_{ij}^A(\bq)+D_{ij}^B(\bq)$,
where \be
D_{ij}^A(\bq)=\frac{1}{N}\sum_{\mu=1}^p\xi_i^\mu\xi_j^\mu\left[\Omega_\mu+\frac{1}{\sqrt{N}}\sum_j\xi_j^\mu
\tanh(\beta q_j)\right]^2 \ee \be
D_{ij}^B(\bq)=\frac{1}{N^2}\sum_{\mu=1}^p \xi_i^\mu\xi_j^\mu
\sum_k (\xi_k^\mu)^2\left[1-\tanh^2(\beta q _k)\right] \ee The
term $D_{ij}^A(\bq)$ mainly describes  fluctuations due to the
randomly drawn indices $\mu(\ell)$; $D_{ij}^B(\bq)$ reflects
mainly the fluctuations generated by decision noise. Comparison
shows that the diffusion matrices given in \cite{GMS2000} and
\cite{MC2001} are both approximations.

Having resolved the temporal regularization, one can now proceed
to  analyze the MG  upon adapting to continuous time each of the
generating functional analysis steps taken for the batch MG models
(upon adding external fields). Since we choose
$\Delta_N=\order(N^{-1})$ the impact of studying
(\ref{eq:masterequation}) rather than (\ref{eq:markovchain}) (i.e.
the uncertainty of our clock) is guaranteed to vanish as
$N\to\infty$. The disorder-averaged generating functional is now a
path integral: \be \overline{Z\left[\bpsi\right]}=
\overline{\room\bigbra e^{-i\sum_i\int\!
dt~\psi_i(t)q_i(t)}\bigket} \label{eq:onlineZ} \ee Accepting the
potential technical problems associated with steepest descent
integration for path-integrals (which here imply additional
assumptions on the smoothness of the correlation- and response
functions), one can adapt the batch procedures, take the limit
$N\to\infty$, and find again an effective `single agent' process:
 \be
\frac{d}{dt}q(t)=\theta(t)-\alpha
\int_0^t\!dt^\prime~R(t,t^\prime)~ \bra
\sigma[q(t^\prime),z]\ket_z+\sqrt{\alpha}~\eta(t)
\label{eq:singleonlineagent} \ee The retarded self-interaction
kernel is given by \be
R(t,t^\prime)=\delta(t-t^\prime)+\sum_{\ell>0}(-1)^\ell
(G^\ell)(t,t^\prime) \label{eq:onlinekernel}
 \ee (i.e. the continuous-time equivalent of
the previous $[\one+G]^{-1}$, written as a power series), whereas
the zero-average Gaussian noise $\eta(t)$ is here characterized by
\begin{eqnarray}
\hspace*{-10mm} \bra\eta(t)\eta(t^\prime)\ket&=& \sum_{\ell
\ell^\prime\geq 0}(-1)^{\ell+\ell^\prime}\int_0^\infty\!\!ds_1
\ldots ds_\ell ds_1^\prime \ldots ds_{\ell^\prime}^\prime
\prod_{uv}\left[1+\frac{1}{2}\eta
~\delta(s_u-s_v^\prime)\right]\nonumber \\ && \hspace*{-15mm}
\times G(s_0,s_2)\ldots
G(s_{\ell-1},s_\ell)[1+C(s_\ell,s_{\ell}^\prime]G^\dag(s_{\ell}^\prime,s_{\ell-1}^\prime)\ldots
G^\dag(s_1^\prime,s_0^\prime) \label{eq:onlinecovariance}
\end{eqnarray}
(which would only for $\eta=0$ be a continuous-time version of
previous expressions).
 The order parameter kernels $C$ and $G$ are to be solved from
 \be
 C(t,t^\prime)=\bigbra \bra
\sigma[q(t),z]\ket_z \bra\sigma[q(t^\prime),z]\ket_z \bigket
\label{eq:onlineC} \ee
\be
 G(t,t^\prime)=
\frac{\delta}{\delta\theta(t^\prime)} \bigbra
\bra\sigma[q(t),z]\ket_z\bigket \label{eq:onlineG} \ee

The closed macroscopic equations
(\ref{eq:singleonlineagent},\ref{eq:onlinekernel},\ref{eq:onlinecovariance},\ref{eq:onlineC},\ref{eq:onlineG})
for the original so-called on-line MG with random external
information (in the limit $N\to\infty$) are claimed to be exact
for any $\alpha$ and any type of decision noise. Although they are
again not easily solved in general, they have clarified many of
the debates and apparent contradictions in previous studies. More
specifically, they reveal that
\begin{itemize}
\item
The ergodic stationary state in the regime $\alpha>\alpha_c(T)$,
which is calculated easily, is {\em  independent} of $\eta$, and
consequently identical to that of the batch MG and the replica
solution in \cite{CMZ2000,MCZ2000}. For $\alpha<\alpha_c(T)$,
however, this is no longer the case: batch and on-line MG dynamics
are now truly different. Previous methods had not yet succeeded in
producing exact macroscopic laws for this regime.
\item
Premature truncation of the Kramers-Moyal expansion after the
diffusion term (leading to a Fokker-Planck equation), provided
with the correct diffusion matrix (i.e. neither of the ones
proposed so far), would for $N\to\infty$ have led to the correct
macroscopic equations. Apparently, the natural order parameters of
the problem are not sensitive to the weak non-Gaussian
fluctuations.
\end{itemize}

\section{Discussion}

Minority Games are a specific class of disordered stochastic
systems without detailed balance.  In this contribution I have
tried to give a (biased) introduction to their definitions,
properties, and non-equilibrium statistical mechanical theories. I
 claim that the generating functional analysis techniques
proposed by De Dominicis \cite{DeDominicis} (which allow for the
derivation of exact, closed macroscopic dynamic equations in the
limit $N\to\infty$) are the canonical tools with which to study
these systems, both (a priori) in view of the specific nature of
the MGs as statistical mechanical systems and (retrospectively) in
view of the insight and clarification which application of these
techniques have already generated. Compared to some of the other
methods in the disordered systems arena, generating functional
analysis is immediately seen to be in principle mathematically
sound. As always in applied areas, one should worry about the
existence of limits and integrals, but these are quite
conventional reservations (in contrast to e.g. replica theory);
the only serious step is the application of saddle-point arguments
in path integrals (for the on-line MG), but, again, this is a
familiar type of worry in statistical mechanics about which much
has already been written.

In addition the generating functional analysis  methods have
opened the door to many interesting new questions which can now
start to be addressed. At the level of increasing realism, one
could re-introduce multiple strategies ($S>2$), increase the
agents' decision range (e.g. to $\sigma\in\{-1,0,1\}$,
representing the options  buy, inactivity, or  sell), inspect
slowly evolving strategy tables, consider the trading of multiple
commodities, introduce capital and trading costs, etc. This is
perhaps not likely to generate fundamentally new mathematical
puzzles, but would make MGs more relevant to those interested in
markets as such, and thereby promote the power of non-equilibrium
statistical mechanics in a wider community.

At the mathematical level one could now assess the precise impact
of replacing the true market memory by random numbers in the
definition of the external information (there is simulation
evidence that for $\alpha<\alpha_c$ there might be non-negligible
differences after all). In replica approaches this would be nearly
impossible (since it renders the model again non-Markovian), but
within the generating functional framework (where one already
finds a non-Markovian effective `single agent' anyway) there
appear to be no fundamental obstacles. Secondly, one could try
(for the batch and on-line versions) to construct or approximate
further solutions of the macroscopic equations in the non-ergodic
regime $\alpha<\alpha_c$. Thirdly, the generating functional
equations reveal that many of the approximations which have been
used in the past to calculate the volatility $\sigma$ boil down to
assuming the existence of non-equilibrium Fluctuation-Dissipation
Theorems in the stationary state. It would be interesting to
investigate more generally the existence and nature  of such
identities relating the correlation and response functions for
models such as the MG, where (in contrast to e.g. spin-glasses)
detailed balance violations are built-in microscopically, rather
than emerging only in the $N\to\infty$ limit. Any new results
concerning such identities would have considerable implications
for a wider area of non-equilibrium systems, including e.g.
non-symmetric neural networks, cellular automata, and immune
system networks.

\section*{Acknowledgements}

It is my pleasure to thank my collaborators Alexander Heimel and
David Sherrington.

\end{document}